\documentclass[conference]{IEEEtran}
\IEEEoverridecommandlockouts
\usepackage{cite}
\usepackage{amsmath,amssymb,amsfonts}
\usepackage{algorithmic}
\usepackage{graphicx}
\usepackage{textcomp}
\usepackage{xcolor}
\usepackage{booktabs, multirow} 
\usepackage{soul}
\usepackage{braket}
\usepackage[T1]{fontenc} 

\newtheorem{theorem}{Theorem}

\def\BibTeX{{\rm B\kern-.05em{\sc i\kern-.025em b}\kern-.08em
    T\kern-.1667em\lower.7ex\hbox{E}\kern-.125emX}}
\begin{document}

\title{Role of Error Syndromes in Teleportation Scheduling\\
\thanks{This research was supported in part by the NSF grant CNS-1955744, NSF-ERC Center for Quantum Networks grant EEC-1941583, and the MURI ARO Grant W911NF2110325}
}

\author{\IEEEauthorblockN{Aparimit Chandra, Filip Rozp{\k{e}}dek, and Don Towsley}
\\ Manning College of Information and Computer Science, University of Massachusetts Amherst
\\[0.1em]
Email: aparimitchan@umass.edu, frozpedek@umass.edu, and towsley@cs.umass.edu
}

\maketitle

\begin{abstract}
Quantum teleportation enables quantum information transmission, but requires distribution of entangled resource states. Unfortunately, decoherence, caused by environmental interference during quantum state storage, can degrade quantum states, leading to entanglement loss in the resource state and reduction of the fidelity of the teleported information. In this work, we investigate the use of error correction and error syndrome information in scheduling teleportation at a quantum network node in the presence of multiple teleportation requests and a finite rate of remote entanglement distribution. Specifically, we focus on the scenario where stored qubits undergo decoherence over time due to imperfect memories. To protect the qubits from the resulting errors, we employ quantum encodings, and the stored qubits undergo repeated error correction, generating error syndromes in each round. These error syndromes can provide additional benefits, as they can be used to calculate qubit-specific error likelihoods, which can then be utilized to make better scheduling decisions. By integrating error correction techniques into the scheduling process, our goal is to minimize errors and decoherence effects, thereby enhancing the fidelity and efficiency of teleportation in a quantum network setting.
\end{abstract}


\section{Introduction}
Quantum teleportation \cite{bennett_teleporting_1993} is a key primitive enabled by quantum communication networks \cite{kimble_quantum_2008}. It allows one to communicate a qubit over a direct, secure, and heralded communication channel, bypassing the risk of losing the qubit to transmission loss \cite{wehner_quantum_2018}. It seems likely that a common task of quantum devices would be to teleport streams of qubits to some other quantum devices. Examples of such a scenario include a 1st or 2nd generation quantum repeater/switch performing bell swaps \cite{munro_inside_2015,azuma2023quantum,muralidharan_optimal_2016} or communication 
between different memory-based quantum processing units in large-scale quantum computers \cite{monroe_large-scale_2014}. To enable quantum teleportation between two parties, an "entangled state" needs to be shared between them, the most common example of one being an EPR pair also known as a Bell pair\cite{bennett_teleporting_1993}. The creation of an EPR pair between two parties is an active area of research \cite{duan_long-distance_2001, jones_design_2016, azuma2023quantum}. However, it is generally impossible to produce remote entanglement in a truly deterministic manner as the link level communication necessitates the use of fragile photons which can act as carriers of quantum information through fiber or free space. This introduces randomness in the best case due to hardware losses and networking delays, or in the worst case due to the inherent probabilistic nature of a photonic bell state measurement\cite{lutkenhaus_bell_1999} on top of the aforementioned delays. The random nature of EPR pair distribution introduces the need for quantum memories at these devices as qubits that need to be teleported (further referred to as request qubits) will need to wait for an EPR pair.

Noise is a major issue with quantum memories as no quantum system can be truly isolated i.e. a qubit in a quantum memory is constantly interacting with the environment, losing the coherence of the information stored in it. The qubits that need to be teleported will be undergoing some noise process while they are waiting for an EPR pair. Previous work \cite{chandra_scheduling_2022} has looked into how the effects of noise depend on which request qubit you choose to pick if there are multiple waiting in the buffer i.e. your scheduling decision. Problems of this structure are studied in performance modeling in the form of queuing theory \cite{harchol-balter_performance_2013}. Quantum Error Correction \cite{shor_scheme_1995} \cite{chatterjee_quantum_2023} is a promising technique that allows one to protect quantum information from the effects of noise. This is done by encoding logical qubits using many physical qubits allowing one to make syndrome measurements that provide information about the noise process that a qubit may have undergone without providing information on the original state. Quantum Error Correction in a memory in experiment can be implemented via repeated rounds of syndrome detection and corrections \cite{cramer_repeated_2016}. Here we consider the scenario when the physical qubits stored in memory undergo dephasing noise \cite{nielsen_quantum_2010} i.e. $Z$ noise and to protect against it, the logical qubits are stored in a quantum error correction code, namely a $3$ qubit repetition code \cite{nielsen_quantum_2010} in the $X$ basis.


 It has been shown that the extracted error syndrome can be used to improve quantum communication performance in certain quantum network settings. The work of~\cite{namiki_role_2016} looked at how repeater chains can incorporate syndrome information in teleportation based error correction to better overcome losses incurred during transmission through lossy links in a one way repeater chain. Similarly, in~\cite{jing2020quantum} syndrome information from entanglement swapping performed on a logical level is used to better deal with operational errors inside two-way repeaters in the context of quantum key distribution. Finally, \cite{fukui2021all,Rozpędek2021,rozpedek_all-photonic_2023} analyze the performance of two way and one way repeater chains with bosonic error correction where the continuous syndrome information is used for post-selecting high quality links, for improving performance of code-concatenated repeaters and for making more informed entanglement swapping decisions.
 
 In this work, we explore how the teleportation fidelity of a system is affected when error correction is employed to protect against memory errors under generation delays and memory noise in continuous time. We quantify the benefits of incorporating the syndrome information in the scheduling decision for the qubits in such a system. We introduce Freshest Qubit First (FQF) as a scheduling policy which keeps track of the syndromes generated by the request qubits and picks the one with the lowest probability of error. We compare it to Youngest Qubit First (YQF), the scheduling discipline known to be optimal when there is no error correction in the system and scheduling is done according to the time spent by a qubit in the system. We prove that FQF is optimal when one needs to teleport a batch of request qubits. We then explore the scenario in which a node is receiving a continuous stream of qubits that it needs to teleport and compare its performance to YQF. Our findings indicate that FQF outperforms YQF with the increase in performance beiong dependent on various system parameters.
\section{System Model}
\label{sec:Sys_model}
We consider a node in a quantum network that is receiving or generating qubits it needs to teleport to another node. We study three scenarios:
\begin{itemize}
    \item The node needs to teleport a batch of qubits that were generated at the same time with some batch size $b$.
    \item The node needs to teleport a steady stream of qubits being generated with rate $\lambda_r$.
    \item The node needs to teleport a steady stream of qubits being generated with rate $\lambda_r$ but each arrival is a batch of qubits of batch size $b$.
\end{itemize}

The node is equipped with an EPR pair generation platform to generate the resource required for teleportation. There are many ways this could be implemented in hardware, for e.g. atomic platforms \cite{cabrillo_creation_1999} using color centers\cite{bernien_heralded_2013} \cite{stolk_metropolitan-scale_2024}, or, trapped ions \cite{krutyanskiy_entanglement_2023}. The key thing to note is most EPR generation platforms are photon-based and this makes the EPR pair generation probabilistic either due to loss in hardware or the inherent randomness in photonic Bell State measurements, therefore we model the time taken for an EPR pair generation as a random process with mean generation time $1/\lambda_e$ or rate $\lambda_e$. 

\subsection{Memory Noise and Error Correction}
\label{mem}
We assume that the physical qubits stored in memory suffer from dephasing noise or Pauli Z noise \cite{nielsen_quantum_2010} given by:
\begin{equation}
\label{deph_cahannel}
    \mathcal{E}_Z(\rho) = (1-p)\rho + pZ\rho Z^\dagger,
\end{equation}

where $p$ is the phase flip probability, i.e., the state is unchanged with probability $1-p$. In a continuous time model $p  = (1-e^{-\Gamma t})/2$ where $\Gamma$ is the decay rate and $t$ is the time spent in the memory. $\Gamma$ is the inverse of what is commonly known as $T_2$ time in literature. We chose to focus on dephasing noise as it is often considered to be the dominant source of noise in many quantum memories. A 3-qubit repetition code is the smallest code that can sufficiently protect against a single phase flip error so we choose that as our logical encoding of the logical memory slots.

As soon as a request qubit arrives, it is encoded in the 3-qubit repetition code and stored in memory as a logical qubit and the system attempts to generate an EPR pair. The stored logical qubit undergoes repeated syndrome measurements and corrections every $\tau$ time steps i.e. the error syndromes are measured using two ancillary qubits and a corresponding error correction operation is applied to the relevant physical qubit. As we wanted to focus on networking delays we consider gates to be perfect fidelity and the only real limitation one has in making $\tau$ as small as possible is gate times. Such a quantum storage system was experimentally demonstrated in \cite{cramer_repeated_2016}.

As soon as an EPR pair becomes available, a request is chosen for teleportation, one last round of measurement detection and decoding is done and the request qubit is teleportated.


\subsection{Scheduling Choices}
\label{sec:Scheduling}
In the case there are multiple request qubits available on an EPR pair's arrival, the system will choose a request qubit to teleport according to a scheduling policy. There is also a choice to be made as to what to do if a qubit arrival occurs to a full memory. In this work we consider three scheduling disciplines with a pushout buffer control mechanism The three scheduling disciplines considered in this work are as follows:

\textbf{Oldest Qubit First}:
Oldest qubit first will make the scheduling choice according to arrival time. It will pick the oldest request in memory to teleport. This scheduling discipline is more commonly known as First in First Out literature \cite{doshi_overload_1986}. The primary benefit of this is that the order of information arrival and departure is the same.

\textbf{Youngest Qubit First}:
Youngest qubit first also will schedule according to arrival time but it will pick the youngest request first. Also known as last in first out or LIFO in classical literature. In \cite{chandra_scheduling_2022} the authors find that in the absence of error correction, this discipline maximizes the fidelity of the teleported information.

\textbf{Freshest Qubit First}:
Freshest qubit first will maintain a priority queue for the qubits in memory with priority calculated using the error likelihood of the logical qubits conditioned on the syndrome histories generated by the qubits. It will pick the one with the lowest error likelihood. This should intuitively be optimal as it linearly searches for the optimal solution. The classical control required for smarter scheduling is an additional engineering challenge.

\subsubsection{Buffer Control}\label{BC}
For the buffer control policy, we adopt a pushout policy. If a request qubit arrives at a full buffer, the system will push out the request qubit with the lowest priority. In the case of OQF and YQF, it would mean discarding the oldest requests in the buffer to make space for the new arrival. FQF would similarly kick out the request with the highest probability of error.

\section{Error Likelihood Analytics}
\label{sec:error_analytics}
Let us consider a qubit that has undergone $n$ rounds of syndrome measurements. A qubit in a 3 qubit repetition code generates 4 possible 2 bit syndrome measurements $\{(1,1), (0,1), (0,0), (1,0) \}$ the first bit $S_0$ corresponds to the stabilizer measurement $\{X_1X_2\}$ and the second bit $S_1$ corresponds to the stabilizer measurement $\{X_2X_3\}$. Given $p$ or the phase flip probability in a round it is possible to calculate the likelihood error after $n$ rounds given the syndrome history. 

After every round of syndrome detection, there is a certain probability that the correction was incorrect and we have logically flipped the stored qubit. This is because if the more than one qubit flips, one cannot identify the error and the syndrome will either not detect it in the case of all three qubits flipping, or it will trigger the syndrome measurement which would have occurred if the error was on the unflipped qubit causing the correction operation to logically flip the qubit. Eitherway, at every syndrome measurement there is a probability that you might have a logical error at the end of it. 

Let $Pr[e]$ denote the probability of logical error qubit and $Pr[e']$ denote the probability of no error. Then $Pr[e|(S_0, S_1)]$ is the probability that there is an error after a round of error correction conditioned on the fact that the syndrome bits for that round were $S_0$ and $S_1$. For a physical phase flip probability $p$

\begin{equation}
\label{e+1}
    Pr[e|(0, 0)] = \frac{p^3}{(1-p)^3 + p^3},
\end{equation}

i.e. we are in the case that all three qubits flipped and,

\begin{equation}
    \begin{split}
        \label{e-1}
        Pr[e|(0, 1)] &= Pr[e|(1, 0)] = Pr[e|(1, 1)]\\
        & = \frac{p^2(1-p)}{p(1-p)^2 + p^2(1-p)} = p,
\end{split}
\end{equation}

i.e. we are in the case that 2 qubits flipped. Since $(0,1), (1,1), (1,0)$ are statistically identical for this analysis, we will label these $``-1"$ i.e. saying that we got a syndrome measurement $``-1"$ means that the syndrome measurement was one of $(0,1), (1,1), (1,0)$. Similarly, we label $(0, 0)$ measurement as $``+1"$.

We can model the number of logical flips conditioned on the number and kinds of measurements as binomial random variables. Let $M_{+1}^a$ be the random variable for the number of logical flips suffered by a qubit given $a$, $``+1"$ measurements, similarly let $M_{-1}^b$ be the random variable for the number of logical flips suffered by a qubit given $b$, $``-1"$ measurements. We can model $M_{+1}^a \sim Binom(a, Pr[e|``+1"])$ and $M_{-1}^a \sim Binom(a, Pr[e|``-1"])$. 

Given these random variables, we can calculate our entity of interest $Pr[e'|\hat{S}]$, the probability of no logical error conditioned on syndrome history $\hat{S} = [a, b]$ where $a$ is the number of $``+1"$ measurements and $b$ is the number of $``-1"$ measurements. Since two logical flips negate each other,

\begin{equation}
\label{pr_e}
    \begin{split}
        &Pr[e'|\hat{S} = [a, b]] = Pr[M_{+1}^a +M_{-1}^b \text{is even}]\\
        &= \frac{1 + ((1-2Pr[e|``+1"])^{a} (1- 2Pr[e|``-1"])^b}{2}.
    \end{split}
\end{equation}

FQF will uses \eqref{pr_e} to schedule the next teleportation as it will utilise this information to schedule the qubit with the smallest error likelihood.

\subsection{Teleportation Fidelity}
To assign physical meaning to the success likelihood calculated in \eqref{pr_e}, we fix our request qubits to be $\ket{+}$ states i.e. the worst case scenario for a dephasing channel. The repeated error correction process described in section \ref{mem} acting on a $\ket{+}$ state can be described as a dephasing channel after decoding:
\begin{equation}
\label{ec_channel}
    \begin{split}
        \mathcal{E}(\rho_+, \hat{S}) &= Pr[e'|\hat{S}] \rho_+\\
        &+ (1-  Pr[e'|\hat{S}]) \rho_-.
    \end{split}
\end{equation}
Here $\hat{S}$ is the syndrome vector containing the counts of the syndrome results, $\rho_+ =\ket{+}\bra{+}$, and $\rho_- =Z\ket{+}\bra{+}Z^\dagger = \ket{-}\bra{-}$.

Since quantum teleportation is a linear channel, the teleported state is equivalent to the right hand side of \eqref{ec_channel} this state so now we can calculate the fidelity of the teleported state to the initial state $\ket{+}$. Fidelity of a pure state to some mixed state $\rho$ is given by:
\begin{equation}
    F(\rho, \ket{\psi}) = \bra{\psi}\rho\ket{\psi}.
\end{equation}

Applying \eqref{ec_channel} yields:
\begin{equation}
\begin{split}
   F(\mathcal{E}(\rho_+, \hat{S}), \ket{+}) &= Pr[e'|\hat{S}]\bra{+}\rho_+\ket{+} \\
    &+ (1- Pr[e'|\hat{S}])\bra{+}\rho_-\ket{+} \\
    &= Pr[e'|\hat{S} = (a, b)].
\end{split}
\end{equation}

To evaluate our policies, we calculate $F(\mathcal{E}(\rho_+, \hat{S}), \ket{+})$ for the teleported qubits as our figure of merit.

\section{Batch Arrivals}

We first consider the scenario of teleporting a batch of qubits that are generated simultaneously. One example of such a system would be teleportation of a higher level block of some other encoding like a loss protection encoding block described in \cite{namiki_role_2016}. We focus on this scenario as in this setting all the qubits are generated at the same time so timing based scheduling policies such as OQF and YQF are reduced to random selection. To be precise, we are not considering a continuous arrival of batches rather focusing on a single batch. In this case, wee have

\begin{theorem}
\label{theorem}
Of all service policies, FQF minimizes probability of logical error for serving arrivals within a single batch.
\end{theorem}
The proof can be found in the appendix.



\section{Evaluation through simulation}
\label{sec:Evaluation}



We evaluate the scenarios described in Section \ref{sec:Sys_model} through a continuous time event based simulation. The following are the parameters that we consider:
\begin{itemize}
    \item $\tau$: time between two error correction rounds,
    \item $T_2$: dephasing time
    \item $\Gamma = 1/T_2$: phase flip rate,
    \item $\lambda_r$: request arrival rate,
    \item $\lambda_e$: EPR generation rate,
    \item $b$: batch size,
    \item $B_r$: request buffer size,
    \item $\rho$ : load or traffic intensity.
\end{itemize}

\subsection{Batch teleportation}
Here we quantify the benefit obtained from using FQF (Theorem \ref{theorem}). We evaluate a setting where multiple request qubits are generated at the same time and the EPR pairs are generated on demand according to a Poisson process with rate $\lambda_e$. The system parameters are chosen on the basis of the experiment reported in \cite{cramer_repeated_2016}.

\begin{figure}
    \centering
    \includegraphics[scale = 0.6]{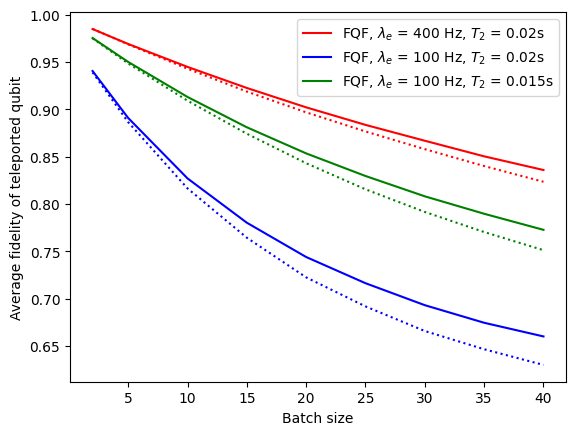}
    \caption{Average $Pr[e']$ for different batch sizes and $\tau = 0.003 s$ under YQF and FQF, dotted lines correspond to YQF.}
    \label{fig:batch}
\end{figure}

\begin{itemize}
    \item $\tau = 0.003 s$,
    \item $\Gamma = 50Hz$.
\end{itemize}
As stated previously, in this setting $OQF$ and $YQF$ are equivalent to random selection. We plot the average teleportation fidelity of the batch in Fig. \ref{fig:batch} for both YQF and FQF for two different rates. We observe that increasing the batch size increases the benefit gained by FQF, This intuitively makes sense as the larger the batch size, the more opportunities there are for random selection to be sub-optimal. We also observe the fact that the difference decreases at higher rates. This follows from the fact that at higher rates, the number of rounds of error correction between two teleportations decreases. 


Let $N$ be a random variable that denotes the number of error correction rounds between two consecutive teleportations in the system. As $N$ increases, the differences in the syndrome histories of the stored request qubits also increase and decisions made by FQF will increasingly differ from those made through random selection (YQF). Last, note that $N$ increases as the EPR generation rate $\lambda_e$ decreases, thus explaining the observed trend in Fig. \ref{fig:batch}.

Another observation from the graph is that the advantage of FQF over YQF increases at lower coherence times. This is because it increases the fidelity loss from a suboptimal choice while also increasing the likelihood of YQF making a suboptimal choice. This can be captured by the variance of the random variable $G$ where $G$ is the number of "$+1$"s generated in $N$ syndrome rounds. Note that $G$ can be modeled as a binomial random variable $G \sim Binom(N, p^3 + (1-p)^3)$ where the number of trials is the number of error correction rounds and the success probability is the probability of getting a "$+1$" measurement. The higher the variance of $G$, the higher the difference between an optimal and suboptimal choice.

\subsection{Continuous Arrivals}
\begin{figure}[t]
\includegraphics[scale = 0.6]{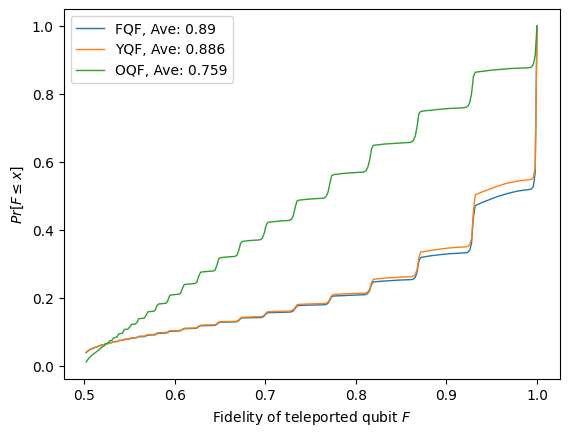}
\caption{CDF of $Pr[e']$ in an infinite buffer system with on demand EPR generation where $\tau = 0.003 s, \Gamma = (1/0.02) Hz = 50Hz, \lambda_r = 90 Hz, \lambda_e = 100 Hz$}
\label{error_prob_underload}
\end{figure}
We move on to evaluate a scenario where the system receives a steady stream of request qubits arriving according to a Poisson process with rate $\lambda_r$. We consider a node with infinite memory for the request qubits and an on-demand EPR generation platform i.e. EPR pairs are generated only when there are requests in the system. 
In Figure \ref{error_prob_underload}, we plot the cumulative distribution function of teleportation fidelity of a random request qubit in the system. We observe that FQF performs better than YQF but not by a significant amount. Of course, OQF performs the worst as expected. The stepwise nature of the CDF can be explained by the fact that the aforementioned parameters lead to an effective physical bit flip probability in a round $p \approx 0.069$ which makes $Pr[e|``+1"] = 0.00042$ and $Pr[e|``-1"] = 0.069$. This means that the likelihood of error is affected more by the number of "$-1$" syndromes causing the step, with the number of "$+1$"s having minimal effect on the error probability. As we can see the probability steps occur around $Pr[e'| \hat{S} = (0, x)]$ where $x \in \{0, 1, 2, ...\}$, e.g. $Pr[e'|a= 0, b = 0] = 1$, $Pr[e'|a= 0, b = 1] = 0.93$, $Pr[e'|a= 0, b = 2] = 0.87$ where $b$ is the number of $-1$s. Each step indicates the number of $"-1"$ syndromes accumulated by the request with the difference in the height of the jump signifying the probability of one less many $"-1"$s as it is a CDF.

\subsection{Effects of Continuous Batch arrivals}
Lastly in Fig. 3 we evaluate the scenario where requests arrive in fixed size batches ($b = 5$) according to a Poisson process, i.e. at each arrival event, a fixed number of request qubits arrive instead of a single one. We also introduce a fixed buffer size for requests with the respective scheduling policies' buffer control policy as described in \ref{BC}. The first observation here is that the benefit of FQF increases with increasing load. This is an artifact of the pushout decision coming into play. At higher loads in finite buffers, a timing-based policy may sometimes pushout a qubit that would have resulted in a higher fidelity explaining the increase with higher loads. To study the effects of the combination of pushout and batch arrivals, we scale the EPR generation rate and buffer size according to the batch size. The system with $\lambda_e = 100, b = 4, B_r =20$ is the scaled version of $\lambda_e = 25, b = 1, B_r =5$ with respect to batch size. We observe that the batch arrivals increase the fidelity advantage gained. This is consistent with the fact that the randomness introduced in YQFs decisions is due to multiple requests arriving at the same time. This effect is also amplified at higher loads as the pushout decisions of YQF become increasingly random, allowing FQF to perform even better. Lastly, we see that in the absence of batch arrivals, the higher EPR rates do not affect the performance advantage gained by FQF if load is kept constant. This stems from the fact that the throughput (amount of pushout) is dependent on the load and the buffer size.

\begin{figure}
    \centering
    \includegraphics[scale = 0.6]{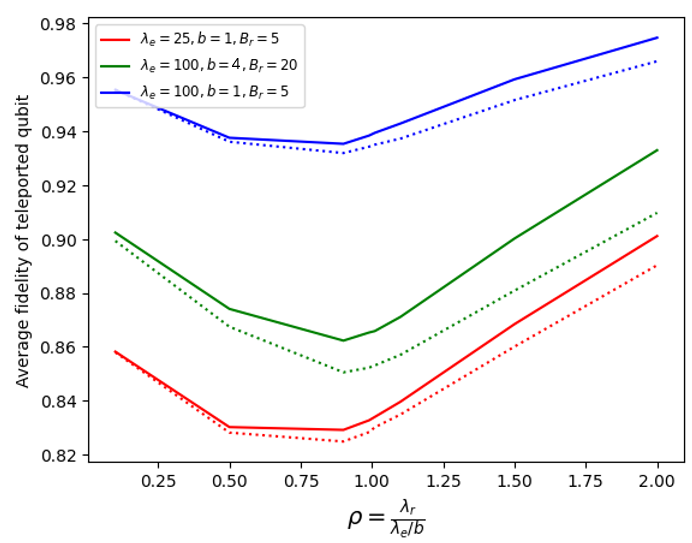}
    \caption{Average fidelity of teleportation with changing load for different rates and buffer sizes. Solid lines are for performance under FQF and dotted lines are for YQF. The EPR generation rates and buffer sizes are scaled according to the batch size with $\lambda_r$ adjusted to keep the load on the system consistent with the x axis.}
    \label{fig:enter-label}
\end{figure}

\section{Conclusion}

We showed that syndrome information can be utilised to increase the fidelity of teleported qubits at a quantum network node. We introduce the queuing discipline Freshest Qubit First and compare its performance to timing based scheduling schemes by evaluating its performance on near term hardware systems.
We prove that $FQF$ is optimal for scheduling teleportations within a batch of qubits and show that the benefit increases with increasing batch sizes. In settings with continuous arrivals, FQF maintains its advantage with increased performance at higher traffic intensity due to smarter pushout decisions. In a combined scenario, the increase in performance is amplified as FQF is able to optimize both batch scheduling and pushout decisions. 

\subsection{Future Work}
In Section \ref{sec:Evaluation} we conjecture about the difference in performance of FQF and YQF and hypothesize that it depends on the variance of the distribution over the the syndrome strings (number of $+1$ and $-1$ measurements) generated before a teleportation which in turn depends on the time scales of error correction, teleportation rate and decay rate. It is possible to calculate this variance and analytically evaluate the performance of YQF. It might be possible to exploit this fact to find analytical bounds for the performance of FQF as YQF seems to lowerbound FQF.
Another open question that arises is the effect of larger codes as larger codes turn the binomial distributions into multinomial distributions where each possible syndrome measurement becomes a possible outcome and is weighted by its likelihood of occuring. This should increase the variance of the syndrome strings generated in error correction further increasing the performance gained by FQF.
The storage of EPR pairs in such a manner is also an interesting problem, as the error probability depends on the syndrome measurements of two qubits in spatially separated nodes. FQF can be applied very easily if one has full knowledge of the syndromes on the two qubits, but if for some reason one only has partial knowledge pushout control becomes quite challenging, especially if one considers repeater chains.
Lastly, we plan on seeing if it is possible to extend Theorem \ref{theorem} to account for more than just batch arrivals and extend it to the case of continuous arrivals.

\bibliographystyle{unsrt}
\bibliography{refs}
\vspace{12pt}
\section{Appendix}
\subsection{Proof of Theorem 1}
Consider $k$ request qbits stored in memory at same time. Rounds of syndrome measurements are computed every $\tau$ timesteps. 
Let $m_{it}$ denote number of $-1s$ for qubit $i = 1, ..., k$ at time $t$. Let $n_{it}$ denote total the total number of syndrome rounds at time $t$. $n_{it} = \rfloor t/\tau \lfloor$
Assume EPR pairs are generated at times $0< t_1 < t_2 < ... <t_k$. Let FQF be policy that serves qubit with smallest $m_{it}$.
Let $\pi$ be some arbitrary policy. Suppose $\pi$ does not pick the freshest qubit at time $t_j, j<k$.
More Precisely suppose there are 2 qubits at $t_j$, $q_1$ and $q_2$, where $q_1$ has $m_1$ "$-1$"s, and $q_2$ has $m_2$ "$-1$"s and $m_2 > m_1$. Suppose $\pi$ does not serve the freshest qubit first; it serves $q_2$ at $t_j$. Suppose $\pi$ serves $q_1$ at $t_l: t_l > t_j$ at which time it has $m_{1}'$ "$-1$"s. 
We create a policy $\pi'$ that behaves exactly like $\pi$ except at $t_j$, $\pi'$ serves $q_1$ and at $t_l$ it serves $q_2$. Under $\pi'$, at $t_l$ $q_2$ will have $m_2 + (m_1'-m_1)$ "$-1$"s, i.e. the number of $-1s$ $q_1$ would have accumulated under $\pi$ are now accumulated by $q_2$.
Last, let $n_j$ be total number of syndromes at $t_j$ and $n_l$ at $t_l$. Let $P_s(m, n)$ be success prob given $n$ syndromes of which $m$ are $-1s$.
Let $p_1 = Pr[e| -1] = p$ and $p_2 = Pr[e| +1 = 0] = \frac{p^3}{p^3 + (1-p)^3}$.
If we restrict ourselves to the regimes where error correction is relevant i.e. $p < 0.5$, we have $p_1 > p_2$.\\
Define $a = 1-2p_1$, $b = 1 - 2p_2$. Since $p_1 > p_2$, $b > a$ Also if $p_1, p_2 > 0.5$ we would not be in a regime where the code is useful, this implicitly gives us the constraint $0 < a, b < 1$.
To prove theorem \ref{theorem}, it suffices to prove:
\begin{equation}
\label{arg}
\begin{split}
    P_s(m_1, n_j) + P_s(m_2 + (m_{1}'-m_1, n_l)\\ - P_s(m_2, n_j) - P_s(m_{1}', n_l) > 0
\end{split}
\end{equation}
as \eqref{arg} can be applied recursively to improve on $\pi'$ and we see that the optimal policy that we arrive at the end is indeed FQF.

Using \eqref{pr_e} we have:
\begin{equation}
    \begin{split}
        &P_s(m_1, n_j) + P_s(m_2 + (m_{1}'-m_1, n_l) \\ &-P_s(m_2, n_j) - P_s(m_{1}', n_l)\\
        &= \frac{1 + a^{m_1}b^{n_j - m_1}}{2} + \frac{1 + a^{m_2 + m_1' - m_1}b^{n_l - m_2 - m_1 + m_1}}{2} \\
        &- \frac{1+a^{m_2}b^{n_j-m_2}}{2} - \frac{1+a^{m_1'}b^{n_l-m_1'}}{2} 
    \end{split}
\end{equation}

Therefore, we have to prove
\begin{equation}
    \begin{split}
        &\frac{1 + a^{m_1}b^{n_j - m_1}}{2} + \frac{1 + a^{m_2 + m_1' - m_1}b^{n_l - m_2 - m_1 + m_1}}{2} \\
        &- \frac{1+a^{m_2}b^{n_j-m_2}}{2} - \frac{1+a^{m_1'}b^{n_l-m_1'}}{2} > 0
    \end{split}
\end{equation}
which after simplification is equivalent to proving:
\begin{equation}
\begin{split}
    &a^{m_1}b^{n_j - m_1} + a^{m_2 + m_1' - m_1}b^{n_l - m_2 - m_1 + m_1}\\ &- a^{m_2}b^{n_j-m_2} - a^{m_1'}b^{n_l-m_1'} > 0
\end{split}
\end{equation}
Factoring,
\begin{equation}
    \begin{split}
        & a^{m_1}b^{n_j - m_1} + a^{m_2 + m_1' - m_1}b^{n_l - m_2 - m_1 + m_1}
        \\ &- a^{m_2}b^{n_j-m_2} - a^{m_1'}b^{n_l-m_1'}\\
        &= a^{m_1}b^{n_j - m_2}[b^{m_2-m_1} - a^{m_2-m_1} \\
        &+a^{m_2 + m_1' -2m_1}b^{n_l-n_j-m_1'+m_1} - a^{m_1' -m_1}b^{n_l-n_j-m_1'+m_2}]\\
        &=a^{m_1}b^{n_j - m_2}[b^{m_2-m_1} - a^{m_2-m_1}\\
        &+ a^{m_1' -m_1}b^{n_l-n_j-m_1'+m_1}[a^{m_2 - m_1} - b^{m_2 - m_1}]]\\
        &= (a^{m_1}b^{n_j - m_2})(b^{m_2-m_1} - a^{m_2-m_1})\\ &\cdot(1 - a^{m_1' -m_1}b^{n_l-n_j-m_1'+m_1})
    \end{split}
\end{equation}
Now it suffices to prove that the 3 factors are positive which we can via contradiction for all of them.
Assuming,
\begin{equation}
(a^{m_1}b^{n_j - m_2}) < 0.
\end{equation}
We get, 
\begin{equation}
\begin{split}
    (a^{m_1}b^{n_j - m_2}) < 0 \iff a^{m_1} < 0 \land b^{n_j - m_2} > 0\\ \lor a^{m_1} > 0 \land b^{n_j - m_2} < 0
\end{split}
\end{equation}
Since $a > 0, m_1 >0, b > 0$, and $n_j > m_2$, 
\begin{equation}
\begin{split}
    (a^{m_1}b^{n_j - m_2}) < 0 \iff False\\
    \therefore (a^{m_1}b^{n_j - m_2}) > 0
\end{split}
\end{equation}
Similarly
$$\because b > a~\&~m_2 - m_1 > 0$$
\begin{equation}
    \therefore (b^{m_2 - m_1} - a^{m_2 - m_1}) > 0.
\end{equation}
Focusing on the last factor. Since you cannot have more $-1$s than the number of checks,
$$\because n_l -n_j \geq  m_1' - m_1,  n_l - n_j - m_1' +m_1 >0$$
and $$\because b < 1, b^{ n_l - n_j - m_1' +m_1} < 1$$
and $$\because m_1' - m_1 > 0~\&~ a < 1\therefore a^{m_1'-m_1} < 1$$
\begin{equation}
    \therefore 1 - a^{m_1'-m_1}b^{ n_l - n_j - m_1' +m_1} > 0
\end{equation}
Since all the factors are positive, their product will also be positive i.e.
\begin{equation}
\begin{split}
    (a^{m_1}b^{n_j - m_2})(b^{m_2-m_1} - a^{m_2-m_1})\\(1 - a^{m_1' -m_1}b^{n_l-n_j-m_1'+m_1}) > 0
    \end{split}
\end{equation}
\begin{equation}
\begin{split}
    \therefore P_s(m_1, n_j) + P_s(m_2 + (m_{1}'-m_1, n_l) \\
    > P_s(m_2, n_j) + P_s(m_{1}', n_l)
\end{split}
\end{equation}
\end{document}